\documentclass[twocolumn,prl,twocolumn]{revtex4}
\usepackage[latin9]{inputenc}
\usepackage{color}
\usepackage{amsmath}
\usepackage{amssymb}
\usepackage{graphicx}
\usepackage{esint}
\usepackage[unicode=true,
 bookmarks=true,bookmarksnumbered=false,bookmarksopen=false,
 breaklinks=false,pdfborder={0 0 1},backref=false,colorlinks=true]
 {hyperref}
\hypersetup{
 linkcolor=magenta, urlcolor=blue, citecolor=blue, pdfstartview={FitH}, hyperfootnotes=false, unicode=true}

\makeatletter
\@ifundefined{textcolor}{}
{%
 \definecolor{BLACK}{gray}{0}
 \definecolor{WHITE}{gray}{1}
 \definecolor{RED}{rgb}{1,0,0}
 \definecolor{GREEN}{rgb}{0,1,0}
 \definecolor{BLUE}{rgb}{0,0,1}
 \definecolor{CYAN}{cmyk}{1,0,0,0}
 \definecolor{MAGENTA}{cmyk}{0,1,0,0}
 \definecolor{YELLOW}{cmyk}{0,0,1,0}
}

\usepackage{amsfonts}\usepackage{tabularx}\usepackage{dcolumn}\usepackage{bm}

\usepackage[outdir=./]{epstopdf}
\usepackage{bbold}
\hypersetup{urlcolor=blue}

\makeatother

\begin{document}

\title{Logarithmic entanglement lightcone in many-body localized systems
}

\author{Dong-Ling Deng, Xiaopeng Li, J. H. Pixley, Yang-Le Wu, and S. Das Sarma}

\affiliation{Condensed Matter Theory Center and Joint Quantum Institute, Department
of Physics, University of Maryland, College Park, MD 20742-4111, USA}

\begin{abstract} 
We theoretically study the response of a many-body localized system to a 
local quench from a quantum information perspective. We find that the local quench 
triggers entanglement growth throughout the whole system, giving rise to a 
logarithmic lightcone. This saturates the modified Lieb-Robinson bound for 
quantum information propagation in many-body localized systems previously 
conjectured based on the existence of  local integrals of motion. In addition, near the localization-delocalization transition, we find that the final states after the local quench exhibit volume-law entanglement.  
We also show that the local quench induces a deterministic 
orthogonality catastrophe for highly-excited eigenstates, where the typical 
wave-function overlap between the pre- and post-quench eigenstates decays {\it 
exponentially} with the system size.
\end{abstract}
\maketitle

\section{Introduction}

It is well known that 
quantum interference induced by the elastic scattering
of random impurities 
can
localize a quantum system.
This leads
to the celebrated Anderson insulator with an absence of diffusion and
zero DC conductivity 
\cite{Anderson1958Absence}. Without interactions,
generic elastic disorder 
localizes all single-particle quantum
states in one and two dimensions~\cite{Evers2008Anderson}.
The fate of isolated interacting quantum systems with strong disorder
has attracted considerable recent attention, both theoretical~\cite{Basko2006metal,altman2015universal,Nandkishore2015many,Pal2010Manybody,Huse2014Phenomenology,Imbrie2016many,Potter2015Universal,Vosk2015Theory,Li2015Many,Li2016Quantum}
and experimental~\cite{Schreiber2015observation,Smith2015Many,Choi2016exploring,Bordia2016Coupling},
under the rubric of many-body localization (MBL).
MBL phases have a number of remarkable properties, ranging from the violation
of the eigenstate thermalization hypothesis~\cite{Deutsch1991Quantum,Srednicki1994Chaos,Rigol2008Thermalization}
to emergent integrability~\cite{Huse2014Phenomenology,Serbyn2013Local,Rademaker2016Explicit,Chandran2014Manybodylocalization,Ros2015Integrals}
and localization-protected orders that are normally forbidden by the
Peierls-Mermin-Wagner theorem~\cite{Vosk2014Dynamical,Huse2013Localization,Bauer2013Area,Bahri2015Localization,Chandran2014Manybodylocalization,Kjall2014Many}.
Most interestingly, MBL systems are not self-thermalizing, which calls into question the generic applicability of quantum statistical mechanics to isolated systems without external baths. In addition, MBL systems are argued, via a variant of the Lieb-Robinson
bound~\cite{Lieb1972Finite}, to exhibit a logarithmic lightcone (i.e., information can
propagate at most logarithmically in time)~\cite{Kim2014Local,Horssen2015Dynamics}, 
which is consistent with the logarithmic growth of entanglement following a global quench~\cite{Znidaric2008Many,Bardarson2012Unbounded,Serbyn2013Universal,Nanduri2014Enanglement,Huse2014Phenomenology}.
This is in stark contrast to both thermal many-body systems,
where entanglement grows ballistically~\cite{Kim2013Ballistic,Schachenmayer2013Entanglement},
and generic quantum systems without disorder,
where an emergent linear lightcone can be proven~\cite{Calabrese2006Time,Gabriele2006Gabriele,Eisert2006General}
and even experimentally observed~\cite{Cheneau2012Light,Langen2013Local,Richerme2014Non,Jurcevic2014quasiparticle}.

\begin{figure}
	\includegraphics[width=0.30\textwidth]{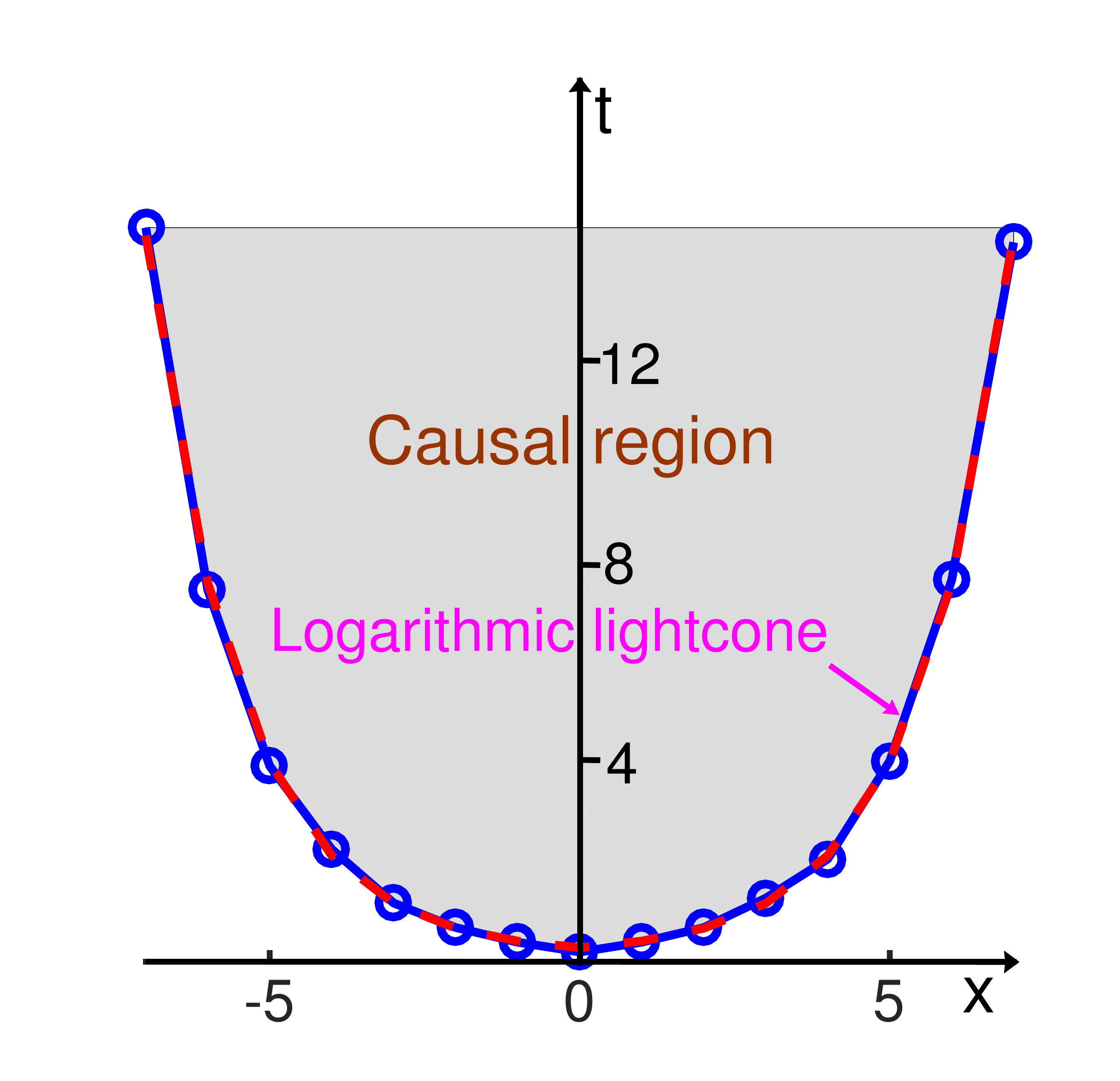}
	
	\caption{Logarithmic lightcone extracted from the local-quench triggered entanglement growth. We begin with highly-excited
		states (obtained by DMRG-X) and calculate $S_{x}(t)$ from TEBD (see text). The
		circled blue  curve is determined by investigating when $S_{x}(t)$
		starts to increase and the red dashed line is a logarithmic fit $x\sim\log t$ to the data. \label{fig:Logarithmic-lightcone} }
\end{figure}

So far, the response of MBL systems to a local quench remains rarely explored in contrast to the global quench which has been studied extensively.
For metallic systems, an arbitrarily weak
local quench can substantially modify the structure of the many-body ground
state.
This leads to the Anderson orthogonality catastrophe (OC)
where the post-quench ground state becomes asymptotically orthogonal to the 
original one~\cite{Anderson1967Infrared,mahan2000many}.
Very recently, a new idea of statistical OC was introduced \cite{Khemani2014nonlocal}
and an exponential-, rather than algebraic-, law has been established
for the ground states of both single-particle (without interaction)
and many-body (with interaction) localized systems \cite{Deng2015Exponential}. Yet, OC for highly-excited states of a localized system  remains largely unexplored. In addition,
the wavefunction dephasing dynamics following a local quench is 
still poorly understood for highly-excited states, except for the general 
expectation that a local quench cannot spread out energy or particle density.
In this paper we provide  insight into the physics of MBL by investigating OC for highly-excited states and probing the 
entanglement dynamics subsequent to a local quench. We address the 
following questions.
Can a local quench, which only changes an intensive amount of energy, 
fundamentally change the wavefunction of a MBL system?
How does quantum information propagate in a MBL phase?
Can we directly probe the logarithmic lightcone?

We consider
an initial disordered interacting Hamiltonian $H_{I}$ (\emph{all} eigenstates of $H_{I}$ are localized), 
and
add a local quench $V_{0}$ to $H_{I}$ at time $t=0$. 
We find that the typical
overlap between the pre- and post-quench excited states is suppressed {\it exponentially} in the
system size, inducing a deterministic OC in the thermodynamic limit. 
Combining 
extensive exact diagonalization
and density matrix renormalization group (DMRG) calculations, we show
that the local quench causes dephasing and triggers an entanglement growth throughout the system.  By comparing the entanglement entropy dynamics for different bipartitions, we 
show that the local quench propagates quantum information through the chain 
logarithmically in time.
This explicitly demonstrates, from first principles, the existence of the 
logarithmic lightcone (Fig.~\ref{fig:Logarithmic-lightcone}) conjectured in MBL systems, without any assumptions about 
the local integrals of motion of the MBL system~\cite{Kim2014Local}.  Moreover, near the localization-delocalization transition, we find that for highly excited states the local quench can trigger an {\it unbounded} entanglement growth with an asymptotic volume-law scaling in the long time limit. 



\begin{figure}
\includegraphics[width=0.46\textwidth]{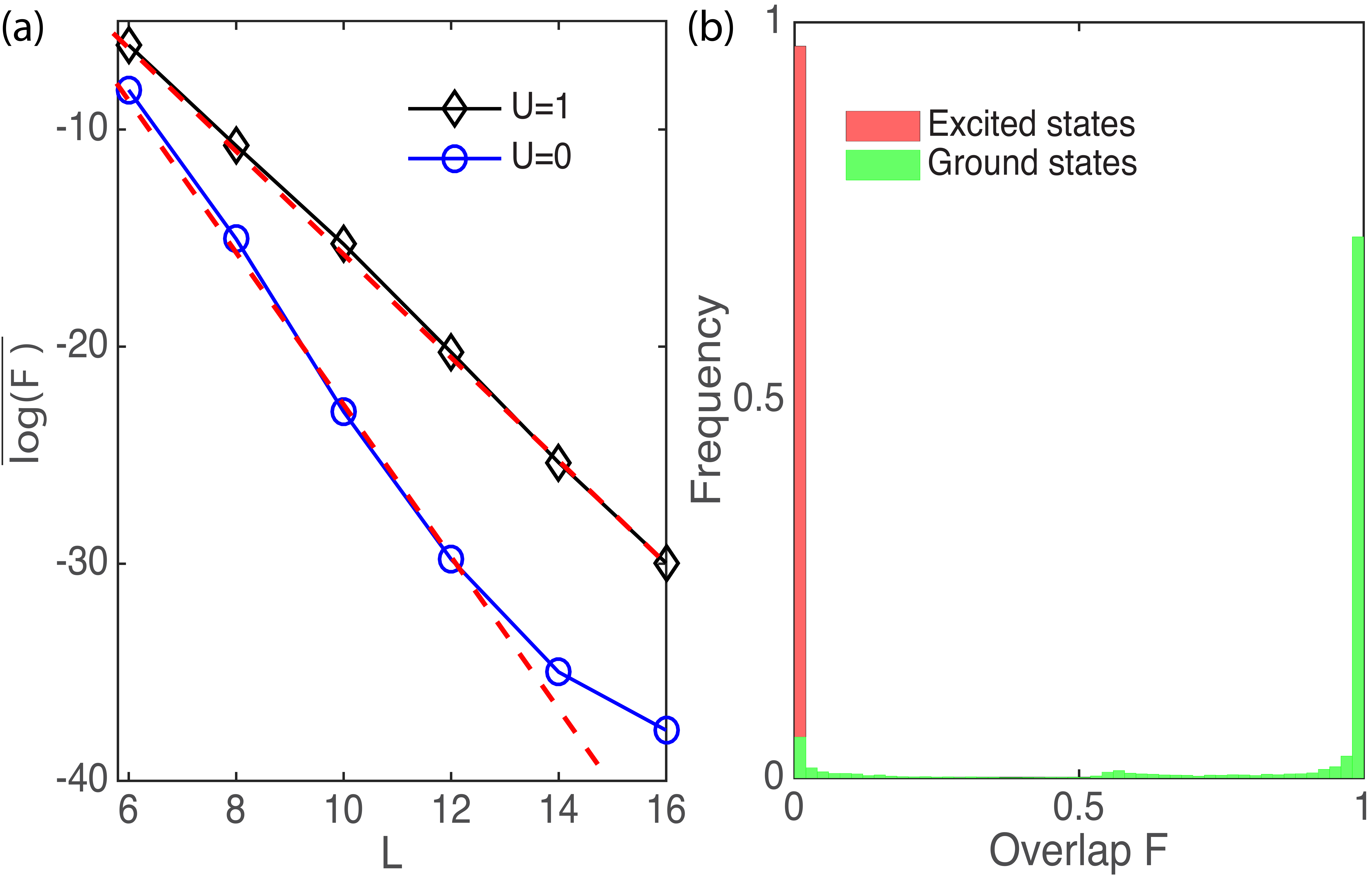}

\caption{Wave-function overlaps between the pre- and postquench eigenstates
in localized systems. (a) Exponential OC of the highly-excited eigenstates,
at $W=15$ and $v_0=6$.
Here $|\psi_{I}\rangle$ and $|\psi_{F}\rangle$ are chosen to be
the excited states in the middle of the spectrum of $H_{I}$ and $H_{F}$,
respectively. The red dashed lines are linear fits to the data. The deviation of the blue curve ($U=0$) from the linear fit at large $L$ is due to the machine precision limit.
 (b) Histogram
of the overlaps of the ground and excited states, respectively,
at $W=15$, $v_{0}=6$, $U=1$ and $L=10$. Unlike
ground state OC which happens stochastically with a probability $\sim\frac{v_{0}}{2W}$
\cite{Deng2015Exponential}, OC for highly-excited state occurs deterministically
(with probability one in the thermal dynamic limit).
\label{fig:Orthogonality-catastrophe-of}}
\end{figure}

\section{Model Hamiltonian}

We consider the generalized interacting spinless one-dimensional (1D) Anderson model 
\cite{Anderson1958Absence}
\begin{eqnarray}
H_{I}  =  \sum_{j=-L/2+1}^{L/2}
-J(c_{j}^{\dagger}c_{j+1}+h.c.)+\mu_{j}n_{j}+Un_{j}n_{j+1},\label{eq:Hamiltonian}
\end{eqnarray}
where $c_{j}$ ($c_{j}^{\dagger}$) are fermionic anihilation (creation)
operators, $n_{j}=c_{j}^{\dagger}c_{j}$ is the corresponding fermion
number operator, $J$ is the nearest-neighbour hopping strength, 
$U>0$ is the nearest-neighbor interaction, and $\mu_{j}$ is the onsite
disorder potential drawn independently from a uniform random distribution
over $[-W,W]$.
We set $J=1$ as the unit of energy and consider half filling. 
The number of disorder realizations ranges
from $10^{5}$ ($L=6)$ to $10^{3}$ ($L=50$).
The Hamiltonian in
Eq.~\eqref{eq:Hamiltonian} has been explored in the context
of single-particle~\cite{Evers2008Anderson} and many-body localization
\cite{bera2015many,Pal2010Manybody,Li2016Quantum}. In the noninteracting
limit $U=0,$ the system is localized for any finite value of $W$
\cite{Abrahams1979Scaling}. 
For 
$U>0$, 
it is now generally
accepted
that the system has a MBL transition at a finite $W$.
With an interaction
strength $U=1$, the reported numerics suggests a critical disorder
strength $W_{c}\gtrsim 7$~\cite{Pal2010Manybody,Luitz2015Many}.

\section{orthogonality catastrophe for highly-excited states}
 To study the effect of local 
perturbations on the eigenstates, we quench the model at $t=0$
by adding a local perturbation $V_{0}=v_{0}c_{0}^{\dagger}c_{0}$
at site $0$. The final Hamiltonian for $t>0$ is $H_{F}=H_{I}+V_{0}$.
We perform full diagonalization on both $H_I$ and $H_F$ and pair up their 
eigenstates $|\psi_I\rangle$ and $|\psi_F\rangle$ per the eigenvalue ordering.
For each pair, we compute the many-body wavefunction overlap (fidelity) 
$F\equiv|\langle\psi_{I}|\psi_{F}\rangle|$ and study its asymptotic behavior 
as a function of the system size.
As shown in Fig. \ref{fig:Orthogonality-catastrophe-of}(a), we find
that the typical wave-function overlap for highly-excited states 
$F_{typ}\equiv\exp(\overline{\log F})$
decays exponentially with the system size
\begin{eqnarray*}
F_{typ} & \sim & \exp(-\beta L),
\end{eqnarray*}
similar to the case of ground states
\cite{Deng2015Exponential}. 
 Here, $\beta>0$ is related to the localization
length $\xi$ and the strength $v_0$ of the local
quench.
This exponential decay is universal for localized systems,
independent of whether it is single-particle or many-body localized.
It originates from the nonlocal charge transfer process induced
by a local quench in localized systems \cite{Deng2015Exponential}.  The exponential decay can also be understood intuitively from the local-integrals-of-motion description of MBL. Within this description, the many-body eigenstates are product states of the local conserved quantities, which commute with the Hamiltonian and have exponentially small tails on operators far away. The pre- and post-quench eigenstates have many different eigenvalues of the local conserved quantities, giving rise to an exponential suppression of the typical overlap between them.

However, it is worth noting that the OC of the highly-excited
states differs from that of the ground states in a statistical
sense. For ground states, the OC occurs with a probability
proportional to the strength of the perturbation \cite{Deng2015Exponential}. While for highly-excited
states, there will always be a multi-particle rearrangement and we
have a deterministic catastrophe with probability one in the thermodynamic limit $L\rightarrow\infty$ \cite{Khemani2014nonlocal}.
This distinction is clearly visible in Fig. \ref{fig:Orthogonality-catastrophe-of}(b).

\begin{figure}
\includegraphics[width=0.481\textwidth]{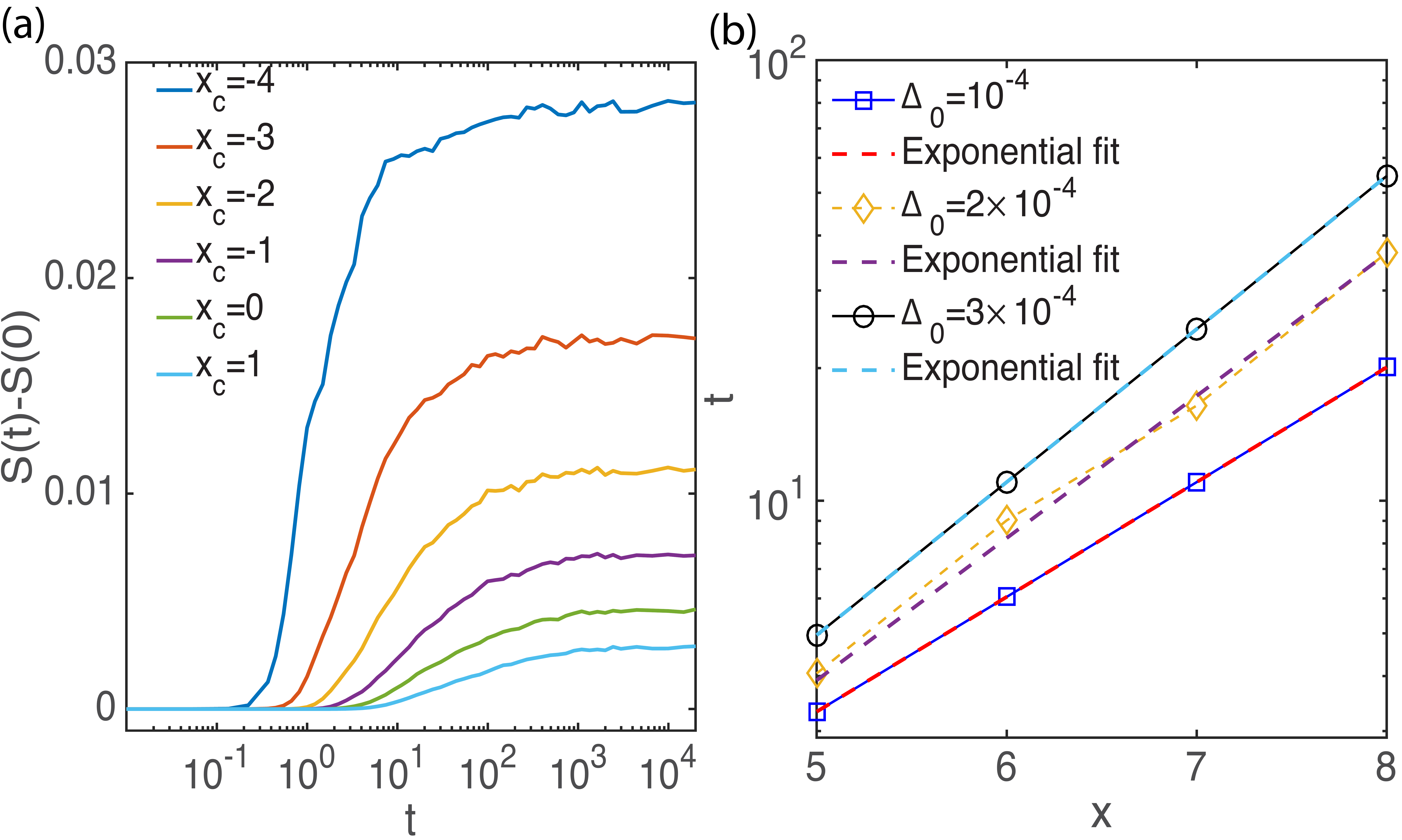}

\caption{Exact diagonalization calculations of entanglement dynamics 
introduced by a local quench and the extracted lightcones for information 
propagation. (a) Deep localized region with $W=15$, $U=1$, $L=12$ and $v_0=6$. We 
begin with a highly-excited state at the middle of the spectrum. $x_{c}$ 
indicates the location where we cut the system into two subsystems A and B 
(cutting the bond between site $x_c$ and $x_c+1$). The local quench is added 
at the left end (site $-5$ in our notation) of the chain. (b) The extracted logarithmic lightcones for different threshold values. For a given threshold value $\Delta_0$, the corresponding logarithmic lightcone is obtained by studying when  $S_x(t)$ starts to increase to pass $\Delta_0$ (see the text). \label{fig:EDEEtCone}
}
\end{figure}

\section{Entanglement growth}

The OC implies that the new eigenstates after the quench are typically 
orthogonal to the corresponding initial ones,
but it does not necessarily imply an essential change in their internal 
properties such as the entanglement entropy.
In a metallic system without disorder, it has been proven recently that
the logarithmic scaling with the system size of the ground state entanglement entropy remains 
intact in spite of the Anderson OC induced by a local perturbation~\cite{Ossipov2014Entanglement}.
In the localized system considered here, notwithstanding the exponential OC, 
the postquench Hamiltonian is still localized and thus its eigenstates still have
area-law entanglement~\cite{Bauer2013Area}.
In neither cases does the local quench change the static entanglement scaling properties 
of the Hamiltonian eigenstates.
However, as we show below, the local quench exerts a strong influence on the 
entanglement dynamics of the highly-excited states in the MBL system.

To study the entanglement dynamics after the quench, we
take an eigenstate $|\psi_{I}\rangle$ of $H_{I}$ and
evolve it under the postquench Hamiltonian $H_{F}$: $|\phi_{F}(t)\rangle=e^{-iH_{F}t}|\psi_{I}\rangle$. We note that this quench is distinct from a quench of the wavefunction \cite{Kjall2014Many}.
We study the time dependence of the von Neumann entropy of 
$|\phi_{F}(t)\rangle$ between two subregions $A$ and $B$:
\begin{equation*}
S(t) = -\text{Tr}[\rho_{A}(t)\log\rho_{A}(t)],
\end{equation*}
where $\rho_{A}(t)=\text{Tr}_{B}[|\phi_{F}(t)\rangle\langle\phi_{F}(t)|]$
is the reduced density matrix of the subsystem $A$. 
To calculate $S(t)$, we use exact diagonalization for small system sizes $L\le 16$. In Fig.~\ref{fig:EDEEtCone} (a), we plot $S(t)$ for different bipartitions for the region deep in the MBL phase ($W=15$).
We find that the entanglement indeed increases, even for bipartitions  far away from the local quench. This can be explained as follows. 
Deep in the MBL phase ($W=15$ and $U=1$), although the wavefunctions are 
localized (product states in the so-called `l-bit' formalism 
\cite{Serbyn2013Local,Huse2014Phenomenology}), the local quench still causes 
dephasing because of the exponentially weak couplings between distant l-bits 
\cite{Serbyn2013Universal} and leads to entanglement growth throughout 
the whole system as shown in Fig. ~\ref{fig:EDEEtCone} (a). This is in 
contrast to the non-interacting Anderson model, where a zero-velocity 
Lieb-Robinson bound~\cite{Hamza2012Dynamical} simply forbids the spread of 
entanglement and restricts the effect of the local quench to within a 
few localization lengths even in the infinite time limit 
\cite{Abdul2016Entanglement}. Interaction enables a spreading of entanglement in the localized system.

\begin{figure}
\includegraphics[width=0.485\textwidth]{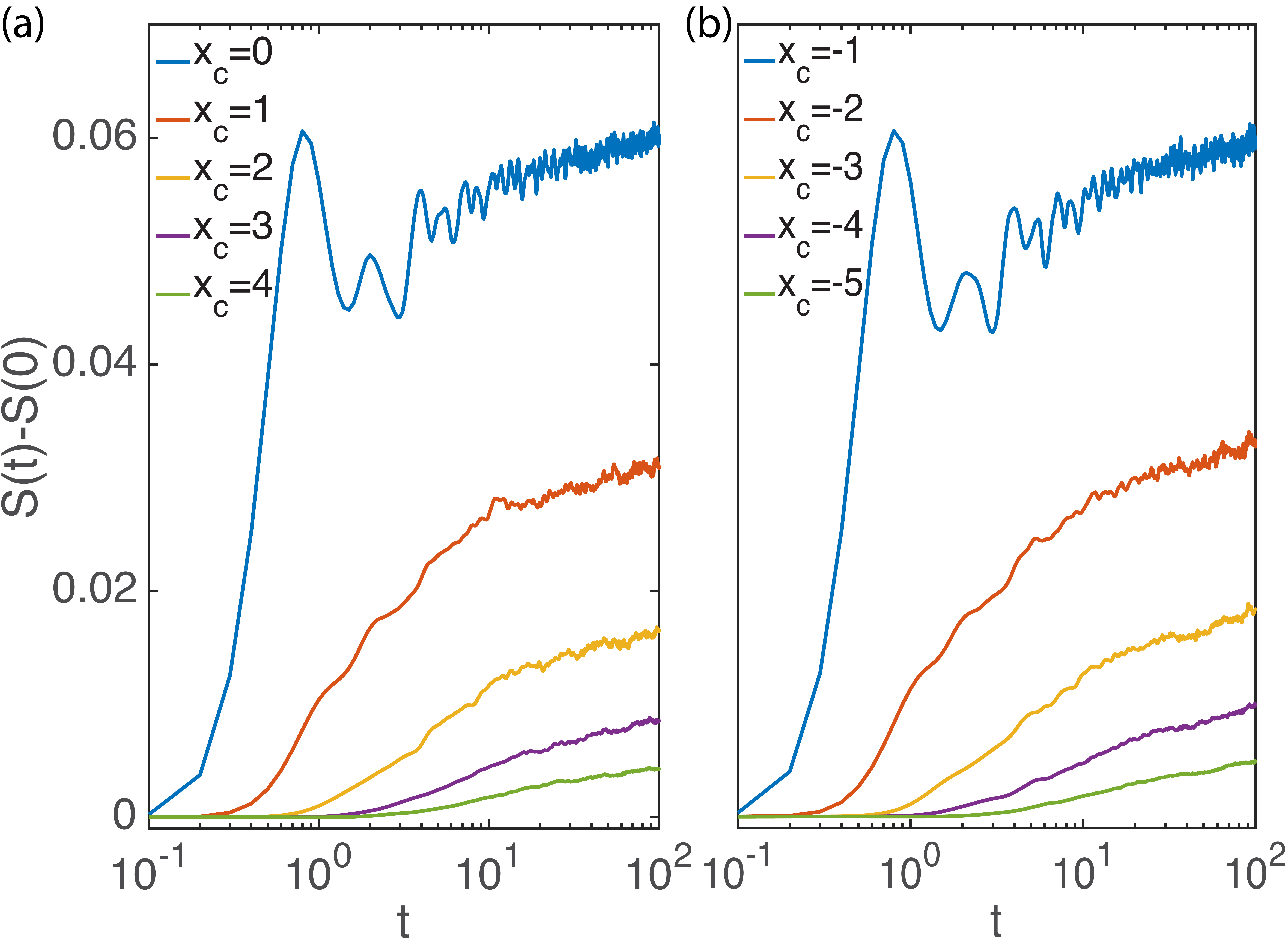}

\caption{The DMRG-X and TEBD calculations of entanglement growth after a local quench starting from a highly-excited eigenstate
$|\psi_{I}\rangle$ of $H_{I}$, for large system sizes $L=50$, with $U=1$, $W=15$, and $v_0=6$.   Here, the local quench is added in the middle of the chain (site $0$). (a) and (b) show the entanglement growth $S(t)-S(0)$ for the right and left half chains, respectively. The logarithmic lightcone plotted in Fig.~\ref{fig:Logarithmic-lightcone} is obtained from the same data shown here.  \label{fig:DMRGEEt}
}
\end{figure}

For larger systems (up to $L=50$), we use 
the recently developed DMRG-X~\cite{Khemani2016Obtaining} to obtain the excited states. The DMRG-X algorithm designed for 1D MBL systems can be seen as a variant of the conventional ground-state DMRG algorithm  based on the matrix product state (MPS) representation ~\cite{Schollwock2005TheDMRG,Schollwock2011Density}. Their procedure is similar, but the key difference is that when we update local matrices, we update each local matrix with a new one that maximizes the overlap with the previous MPS, rather than the one corresponding to the minimal energy.  
We then use the time
evolving block decimation (TEBD) method~\cite{Vidal2003Efficient,Vidal2007Classical}
to perform the time evolution and compute $|\phi_{F}(t)\rangle$ and
$S(t)$. In our calculations, we use the fifth-order Suzuki-Trotter decomposition of the short time propagator and the maximal bond dimension (where we truncate the MPS) is chosen as $\chi_\text{max}=30$. To control the error, we examine the typical variances $\sigma^2=\langle H^2\rangle-\langle H\rangle^2$ (less than $10^{-8}$) in the DMRG-X and check the neglected weight for each truncation ($\sim 10^{-6}$) in the TEBD calculations.

In Fig.~\ref{fig:DMRGEEt}, we plot the entanglement dynamics, calculated by the DMRG-X and TEBD techniques, for different bipartitions with a larger system size $L=50$.  Similar to the exact diagonalization results plotted in Fig.~\ref{fig:EDEEtCone}(a), we again find entanglement growth. 
Another observation is that we can use $S(t)$ as a function of the partition distance from the local quench (Fig.\ref{fig:DMRGEEt}(a)) to provide a rough estimate of the
many-body localization length
$\xi$.
Note that the initial rapid ``boost'' (until
$Jt\sim1$) and ``jitter'' behavior of $S(t)$ arise from the expansion
of wave packets within a distance $\xi$ of the local quench.
However, the wave packet cannot diffuse beyond $\xi$.
The influence of the quench may propagate beyond $\xi$ only via dephasing,
and the entanglement growth after $Jt\gtrsim\xi$ is purely due to 
interactions.
Hence there is no rapid boost or jitter behavior of $S(t)$ for partitions that 
exceed a distance $\xi$ from the local quench as shown by the $x_{c}=2$ curve 
in Fig.~\ref{fig:DMRGEEt}(a).
We therefore conclude 
from Fig.~\ref{fig:DMRGEEt}(a)
that $\xi$ is on the order of a couple of lattice 
spacings for $U=1$.

\section{Scaling of the $t\rightarrow\infty$ entanglement}
In the previous section, we have shown that a local quench can trigger entanglement growth throughout the whole system. We now discuss the post-quench entanglement dynamics in the long time limit. 

\begin{figure}
\includegraphics[width=0.481\textwidth]{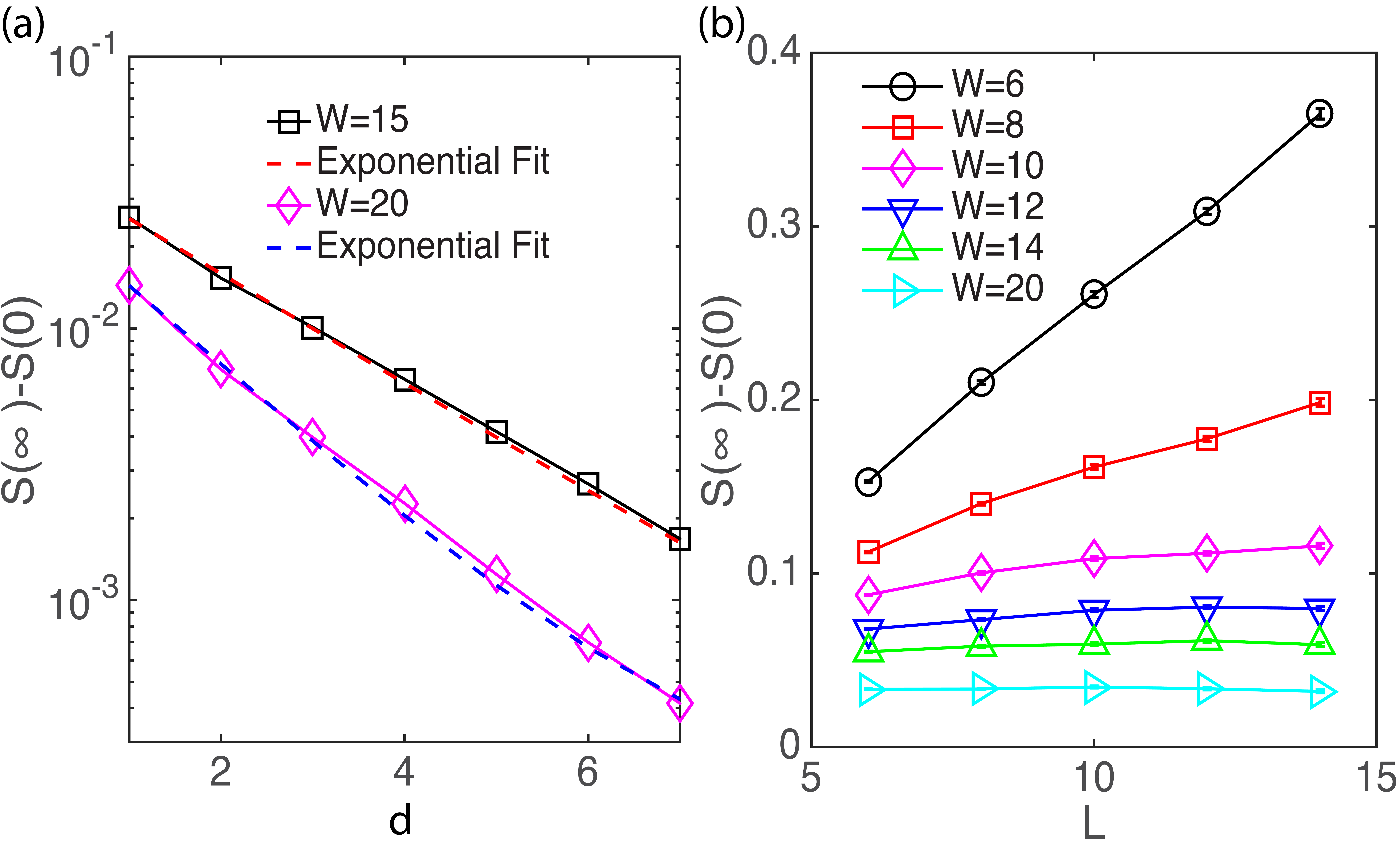}

\caption{ (a) Final entanglement growth versus the distance between the local quench site and the entanglement cut. The results for two different disorder strengths ($W=15,20$) deep in the MBL phases are plotted. In the long-time limit, the net entanglement growths introduced by the local quench decay exponentially with the distance from the quench. Here, the other parameters are $U=1$ and $v_0=6$. (b) Scaling of the saturated entanglement with the system size $L$ for different $W$. Here, we consider a bipartition into two half chains of equal size. Deep in the localized region ($W=14,20$), the final states after the quench exhibit area-law entanglement. In contrast, near the localization-delocalization transition point ($W=6,8$), the local quench gives rise to an unbounded entanglement  growth in the thermodynamic limit $L\rightarrow \infty$. For finite $L$, the saturation value of the entanglement scales linearly with the system size $S(\infty)\sim L$. \label{fig:FinalEE}
}
\end{figure}

In Fig. \ref{fig:FinalEE} (a), we have plotted the saturated (i.e., $t\rightarrow\infty$) entanglement as a function of the distance $d$ between the local quench and the entanglement cut. We find that the saturation values decrease exponentially as the distance increases.
Moreover, as the disorder strength increases, the decrease of the saturation 
value steepens.
This is consistent with the local integrals of motion description of MBL: the $l$-bits have exponential tails on locations far away. We thus expect 
 $S_d(\infty)-S_d(0)\sim e^{-\alpha d/\xi}$, 
where $S_d(t)$ denotes the entanglement entropy for a bipartition where the cut location is $d$ sites away from the local quench, and $\alpha$ depends on the details of the Hamiltonian. When increasing the disorder strength $W$, $\xi$ decreases and thus the net entanglement growth at long time $S_d(\infty)-S_d(0)$ decreases faster. 

In Fig. \ref{fig:FinalEE} (b), the scaling of the saturated entanglement with the system size $L$ is plotted. We find that deep in the localized region, while the local quench indeed induces a finite entanglement growth, the final post-quench states seem to still obey an area-law entanglement. The entanglement growth is finite even in the thermodynamic limit $L\rightarrow\infty$.  This is in sharp contrast to the extensively-studied case of global quench \cite{Znidaric2008Many,Bardarson2012Unbounded,Serbyn2013Universal,Nanduri2014Enanglement,Huse2014Phenomenology}. There, one starts with random product states and an unbounded logarithmic entanglement growth has been shown. In this case, the saturated entanglement follows a volume law even in the parameter regions very deep in the localized side. 
Interestingly, for the local quench studied in this paper, we find that near the MBL transition ($W_{c}\gtrsim 7$~\cite{Pal2010Manybody,Luitz2015Many}), $S(\infty) - S(0)$ does not saturate in $L$ and the local quench can also induce an unbounded entanglement growth when $L\rightarrow\infty$ in the thermal phase ($W=6$). The saturated entanglement satisfies a volume law, as shown in  Fig. \ref{fig:FinalEE} (b) for $W=6$.

\section{Logarithmic lightcone}

We now turn to the important question of how 
fast the influence of the local quench can propagate.
A modified Lieb-Robinson bound $\propto te^{-x}$
has been proposed for MBL systems under certain assumptions~\cite{Kim2014Local}. This bound implies that information can
propagate \emph{at most} logarithmically, i.e. within a logarithmic lightcone.
However, it is not clear whether this bound can be saturated or not in real MBL systems. 
Our local quench procedure provides an ideal setup to study this important bound and directly measure how fast the influence of the quench can spread without making any ad hoc assumption.
To this end,
we first calculate the entanglement dynamics for different 
bipartitions at a distance $x$ from the quench site (denoted by $S_x(t)$),
and then extract $t$
as a function of $x$ by studying when $S_{x}(t)$ begins to increase. In 
numerics, we impose a threshold $\Delta_0$ on 
$\Delta_x(t)\equiv S_x(t)-S_x(0)$ and the function $t(x)$ is determined by the 
time when $\Delta_x(t)$ increases above $\Delta_0$: 
$t(x)=\min\{t|\Delta_x(t)\geq \Delta_0\}$. In the delocalized region ($W<W_c$), 
the influence of the local quench spreads out ballistically, we thus expect a 
linear lightcone. 
In contrast, in the localized region ($W>W_c$), we find a logarithmic 
lightcone $x\sim\log t$, 
as shown in Fig.~\ref{fig:Logarithmic-lightcone} and Fig.~\ref{fig:EDEEtCone}(b).
This {\it saturates} the modified Lieb-Robinson bound in the MBL phase. We stress that our
observation of the logarithmic lightcone comes directly from the numerical data on sufficiently large system sizes, which does not rely on any assumption and is free from any significant finite size effects. 
It establishes that, unlike for
the single-particle localized systems, the Lieb-Robinson bound for
MBL systems cannot be further tightened to a bound constant in time,
consistent with the analytical analysis in Ref.~\cite{Friesdorf2015Local}. Our result is, however, free from any constraining assumptions and is established by numerically studying MBL systems of rather large sizes.

\section{Conclusion}

We have studied the response of a MBL system to a local quench. We establish that a local quench can induce an {\it exponential} OC for highly-excited states in MBL systems, which is deterministic and has important implications on the post-quench entanglement dynamics. Using exact diagonalization and DMRG-X techniques, we demonstrate that in the interacting MBL system local quench triggers an entanglement growth throughout the whole system, in sharp contrast to the noninteracting localized situation where the entanglement is necessarily localized within a localization length. By numerically investigating $S_x(t)$,  we explicitly establish the existence of a logarithmic lightcone conjectured in MBL systems. Our results indicate that quantum information propagates logarithmically in MBL systems and the modified Lieb-Robinson bound cannot be further tightened to a time-independent bound, in contrast to the case of single-particle localization. In addition, we have shown that near the localization-delocalization transition the local quench can trigger an {\it unbounded} entanglement growth with an asymptotic volume-law scaling in the long-time limit.

\section*{ACKNOWLEDGMENTS}
We thank J. Moore and R. Vasseur for stimulating discussions during the early part of this work. We thank R. M. Nandkishore, J. Y. Choi and S. T. Wang for helpful comments, and D. Huse in particular for critical reading of our manuscript and valuable suggestions.  DLD is grateful to Y.-H. Chan, M. L. Wall, and H. -C. Jiang for their help in developing the DMRG code. This work is supported by JQI-NSF-PFC and LPS-MPO-CMTC.   DLD and JHP also  acknowledge additional support from the PFC seed grant ``Thermalization and its breakdown in isolated quantum systems''.
We acknowledge
the University of Maryland supercomputing resources
(http://www.it.umd.edu/hpcc) made available in conducting the research reported in this paper.

\bibliographystyle{apsrev4-1}
\bibliography{Dengbib}

\end{document}